# Relativistic effects on Properties of Halogen Group Elements/Ions

Mohamed Kahil[1,*], Nabil Joudieh[1,*] and Nidal Chamoun[2,3,*]

[1]Department of Physics, Faculty of Sciences, Damascus University, Damascus, Syria
[2]Arabic Language Academy of Damascus, P.O. Box 327, Bld 6, Abdul-Munim Riadh St., Malki, Damascus, Syria
[3]CASP, Antioch Syrian University, Maaret Saidnaya, Damascus, Syria

\* Correspondence: mohamed.kahel@damascusuniversity.edu.sy
nabil.joudieh@damascusuniversity.edu.sy, chamoun@uni-bonn.de.

**Abstract:**

This study investigates the influence of relativistic effects on some atomic properties of the halogen group and gold atoms, including their ions (±1). The analysis covers radii, orbital's energy, first and second ionization energies, electron affinity, and polarizability. The study confirms that the $p_{1/2}$ orbitals contract under relativistic effects, whereas for the $p_{3/2}$ orbitals, the mass-velocity and spin-orbit effects do not appear to cancel each other out completely. This may indicate that the spin-orbit effect grows, when increasing the atomic number, slightly faster than the mass-velocity effect. In addition, expansion of the $np_{3/2}$ orbitals may lead to dilation of the bond length in the related molecules. We found that the non-relativistic Hartree-Fock method gave, for atoms from fluorine to iodine, first ionization energy values with smaller deviations from their experimental ones than other methods involving relativistic and correlation effects. In particular, the method accurately, up to three significative digits, predicts the experimental value for chlorine, and thus can be adopted, discarding other sophisticated methods considering the huge computational effort required by them while not improving much on the agreement with experiment, when evaluating physical/chemical properties of large systems containing light halogen elements. It also predicts an electron affinity of 2.4 eV for the tennessine atom, where it shows also that the relativistic effects play a more important role than in gold atoms.

**Key words**: Gold, Astatine, Tennessine, relativistic and correlation effects, radii, ionization energy, electron affinity, polarizability.

# I- Introduction:

The Group of the 17[th] column of the periodic table, also known as the halogen group, consists of the elements: fluorine(F), chlorine(Cl), bromine(Br), iodine(I), astatine(At), and tennessine(Ts). Among these elements, astatine and tennessine are particularly challenging to study due to their radioactive nature and short-lived isotopes, with half-lives of only 14 and 78 ms for $^{293}$Ts and $^{294}$Ts, respectively. The main difficulty lies in the fact that these elements are only available in extremely small quantities[1]. As a result, a comprehensive related research has been limited on the experimental side. On the other hand, theoretical calculations can provide valuable information for such elements and may sometimes be the only method available to investigate specific aspects of their chemistry.

Relativistic effects can significantly influence various chemical and physical properties of heavier elements. Well-established examples include the yellow color and the nobility of gold. A likely, though not officially proven, consequence is the liquidity of mercury at room temperature[2]. The lead-acid battery generates much of its voltage due to these relativistic effects[3], in that they account for 1-7-1.8 V in a standard 2-V battery cell. Additionally, the compounds cesium gold (CsAu) and rubidium gold (RbAu) behave as semiconductors in the relativistic framework rather than as metals[4]. Because of the relativistic effect, it is the Cs rather than the Fr which appears as the most reactive metal in the periodic table[5].

The effect on the adsorption energy of carbon monoxide on platinum low-index surfaces is so pronounced that the adsorption energy cannot be described by non-relativistic theory based on the Schrodinger equation[6]. However, continuing efforts are required to raise the awareness in the chemistry community of (i) the impact of relativity on chemistry beyond the celebrated consequences, mentioned above, on the gold color and the mercury melting point, and (ii) new developments of relativistic quantum chemical methods allowing for increasingly accurate calculations showing an increasing number of molecular spectroscopic properties. As an example for point (i), a Nature (London) article by Gagliardi and Roos reported calculations for the $U_2$ molecule in which there are three two-electron and four one-electron bonds. Uranium is the heaviest naturally occurring element, and relativistic effects play a major role in $U_2$ and other actinide systems[3].

The relativistic first-order one-electron Hamiltonian h[1] is given by:

$$h^1 = h^{dir} + h^{ind} ; \quad (1)$$

where the direct perturbation h[dir] consists of three major terms. The first one is the mass-velocity correction h[MV] representing the relativistic correction on the kinetic energy. The second one is the spin-orbit coupling h[SO], whereas the Darwin term h[D] represents the third direct relativistic correction added to the electron potential energy[7]. As in the non-relativistic Schrodinger equation case, the analytical solutions of the relativistic Dirac equation are possible only for one electron hydrogen–like systems[8], where the direct relativistic corrections are given by

$$h^{MV} = \frac{-|\hat{p}|^4}{8m^3c^2}, h^{SO} = \frac{1}{4\pi\varepsilon_0}\frac{Ze^2}{2m^2c^2}\frac{\hat{s}.\hat{L}}{|\hat{r}|^3}, h^D = \frac{1}{4\pi\varepsilon_0}\frac{\pi Ze^2}{m^2c^2}\frac{\hbar^2}{2}\delta^3(r) \quad (2)$$

The mass-velocity term has the greatest impact on spherical s-states and gives negative shifts in all cases. The spin-orbit coupling leads to positive as well as negative energy shifts depending on the total momentum number j, whereas the Darwin term always gives a positive contribution[9].

In the case of multi-electron atoms, in addition to the three direct effects, indirect ones h[ind] can appear, leading to changes of the Hartree-Fock potential, due to the relativistic first-order change of the occupied atomic orbits (AO)[7].

Halogen group elements end with a p-type orbital that contains five electrons. When it comes to p electrons, the relativistic mass-velocity effect is similar to that observed in s electrons, but it is much smaller. The spin-orbit effect further divides the six p-shell spin orbitals into two $P_{1/2}$ spinors and four $P_{3/2}$ spinors. The combined impact of the mass-velocity and spin-orbit effects on atoms is such that they nearly cancel each other for $P_{3/2}$ electrons, while they reinforce each other for $P_{1/2}$ electrons. This fact results, for the latter spinors, in contractions and energy stabilization which are approximately equal to those observed in s electrons with the same principal quantum number[10]. It has been known [1,11] that the relativistic effects on the (At, Ts) elements are important and can not be ignored. However, to our knowledge, no quantitative study has been performed to analyze the effects of this fact, and we aim in this work to do just this.

Our paper aims to investigate the role of relativistic effects on all elements of the halogen series, focusing on the heavy ones, astatine and tennessine, which makes this study particularly important. The properties that have been analyzed in order to assess the influence of the relativistic effects include the radii and energies of the outer p-orbitals, first and second ionization potentials, electron affinity, and electrostatic polarizability.

Ionization potential (IP) and electron affinity (EA) are fundamental atomic properties that carry significant implications for the electronic structure of elements and provide insights into their chemical behavior, reactivity, compound formation, and the stability of their chemical bonds[12,13]. Electron affinities (EAs) of alkali and halogen atoms (from F to At) were calculated in 1983 using Hartree-Fock and Dirac-Hartree-Fock methods by Nowak et al[14]. Ionization energies and electron affinities of bromine and heavy halogen atoms up to the super-heavy element 117 were calculated in 2006 by Mitin and Wüllen[11], where a two-component quasi-relativistic Hamiltonian based on spin-dependent effective core potentials was used. In a study performed in 2010 [1], the Multiconfiguration Dirac-Fock (MCDF) method was employed in order to calculate the first five ionization potentials, electron affinities, and radii for the element tennessine and its homologue astatine, and the calculations took also into account the main valence correlation effects, and used an estimation for the contributions of the Breit interaction and quantum electrodynamics (QED) effects. In 2015, values of the first and second ionization potentials, as well as electron affinities, for the super-heavy elements 115-117 were computed by Borschevsky et al[13], who used the Dirac-Coulomb Hamiltonian framework, amended for Breit and QED contributions, and applied the same approach to calculate the ionization potentials and electron affinities for the lighter homologous, Bi, Po, and At. The experimental value of the electron affinity of astatine was reported in 2020 by Leimbach et al. to be 2.41578(7) eV[15]. As to the electric dipole polarizability, which is an important atomic property often used in chemistry as a measure of reactivity[16], it constitutes an important ingredient in many applications ranging from optical phenomena (e.g. dielectric constants, refractive indexes) to blackbody radiation shifts, going through atoms in optical lattices, quantum information, interatomic interactions, polarizable force-field calculations, and atomic scattering to name but a few, its values for neutral atoms, including halogens, were gathered from both theoretical and experimental sides, as compiled in [17].

The halogen elements are important in quasi most of organic/inorganic chemistry, physics and biology. Moreover, the gold element has been extensively studied in this regard, and other works have shown it to have a high electro-negativity similar to that of iodine and so is viewed as a pseudo-halide[18]. About the practical applications concerning the heavy halogen elements, we cite for example that one of astatine's isotopes, $^{211}$At, is remarkably well suited for targeted radionuclide therapy of cancer[15]. Theoretical investigations of atomic properties of superheavy elements are important in assigning these atoms to their places in the periodic table and in predicting their chemical and spectroscopic behavior. Furthermore, predictions of atomic properties assist in future experiments, such as spectroscopic measurements or gas-phase chromatography experiments[13].

Gold is included in our study for comparison purposes, since the corresponding relativistic effects on its properties have been extensively reviewed[19-28]. It is shown in [27] that the "local maximum" of the contraction of the 6s shell in the elements Cs (Z = 55) to Fm (Z = 100) takes place at the gold atom position.

The plan of the paper is as follows. We review in section II the theoretical background of the methods used in the study. In section III, we summarize the computational details, followed in section IV by the results and corresponding discussion. We end up in section V with concluding remarks.

II-Theoretical Background

1. **Dirac-Coulomb Hamiltonian**

The Hamiltonian for the full electron-electron interaction is derived from quantum electrodynamics (QED) [29]. However, this interaction cannot be expressed in a closed form and is accessible only through time-dependent perturbation theory. For practical applications, approximations to the electron-electron interaction are necessary. Here, we consider only interactions at the lowest order. The interaction of one electron with another through the exchange of a photon with frequency ω$_{αγ}$ is [30].

$$\hat{V}_{ij}^{\alpha\beta} = \frac{1}{r_{ij}} - \left[ \frac{\alpha_i \cdot \alpha_j}{r_{ij}} \cdot e^{i \cdot \omega_{\alpha\gamma} \cdot r_{ij}} + (\alpha_i \cdot \vec{V}_i) \cdot (\alpha_j \cdot \vec{V}_j) \cdot \frac{e^{i \cdot \omega_{\alpha\gamma} \cdot r_{ij}} - 1}{\omega_{\alpha\gamma}^2 \cdot r_{ij}} \right], \quad (3)$$

where the $\alpha_i$ denote Dirac matrices at the electron i. Upon using the Coulomb gauge $(\vec{\nabla} \cdot \vec{A} = 0)$, it is often a reasonable approximation to set ω → 0. In this limit, the electron-electron interaction simplifies to the Coulomb-Breit interaction [31].

$$\hat{g}^{Coulomb-Breit}(i,j) = \frac{1}{r_{ij}} - \left[ \frac{\alpha_i \cdot \alpha_j}{r_{ij}} + \frac{1}{2} \cdot (\alpha_i \cdot \vec{V}_i) \cdot (\alpha_j \cdot \vec{V}_j) \cdot r_{ij} \right], \quad (4)$$

The first term is the familiar Coulomb interaction, also found in non-relativistic calculations. The term in brackets represents the Breit interaction. The first component of the Breit interaction, known as the Gaunt interaction (derived by Gaunt in 1929) [32], is a current-current interaction that includes spin-spin, orbit-orbit, and spin-other-orbit contributions. The second component, called the gauge term, accounts for the delay in interaction due to the finite speed of light, and it does not appear in the Feynman gauge. The Coulomb interaction is the lowest-order term in perturbation theory, while the Breit interaction enters at order c$^{-2}$ in the expansion, and it is often omitted in many four-component calculations. All four-component results reported here were obtained using the Dirac program [33], which includes only the non-relativistic Coulomb electron-electron interaction. Thus, the Dirac-Coulomb Hamiltonian used in molecular relativistic calculations in the Born-Oppenheimer reference frame is as follows:

$$\hat{H}_{DC} = \sum_{i=1}^{N} [c \cdot \alpha_i \cdot p_i + \beta_i \cdot m \cdot c^2 + V_{ext}(r_i)] + \sum_{i<j} \frac{1}{r_{ij}} \quad (5)$$

Here, c is the speed of light, $\alpha_i$ and $\beta_i$ are Dirac matrices, $p_i$ is the momentum operator for electron i, m is the electron mass, $V_{ext}(r_i)$ is the external potential due to the nucleus, and $r_{ij}$ is the inter-electron distance. This Hamiltonian governs the four-component Dirac spinor wave functions, accounting for both large and small components that reflect relativistic behavior, and includes a sum over the one-electron Dirac Hamiltonians, the Coulomb electron-electron interaction.

2. **The ZORA Hamiltonian**
   **Classification of methods versus components**

We outline very briefly the theoretical foundations of ZORA method utilized in this work, with its scaled version. There are two main categories of methods: (i) techniques that eliminate the small component, including the zeroth-order regular approximation (ZORA), first-order regular approximation (FORA), and normalized elimination of small components (NESC), and (ii) techniques that employ a unitary transformation to decouple electronic and positronic states, such as the Douglas–Kroll–Hess (DKH) and infinite-order two-component (IOTC) Hamiltonians. The zeroth-order regular approximation (ZORA) to the fully relativistic Hamiltonian has become a popular method in relativistic quantum chemistry [34-46]. ZORA is based on an expansion of the four-component relativistic Hamiltonian using a perturbation parameter dependent on the potential [36, 37]. At its zeroth order, this expansion already introduces relativistic corrections to non-relativistic energy values. Unlike the Breit-Pauli approach [47], the ZORA Hamiltonian remains bounded from below, making it feasible for quasi-variational calculations [48]. In atomic and molecular applications, ZORA and related methods [49-51] provide an accurate depiction of valence and sub-valence electrons in heavier elements [36, 37, 48].

**Brief derivation of the ZORA Hamiltonian**

The derivation of the ZORA Hamiltonian begins with separating the electronic and positronic components of the Dirac equation through a unitary transformation known as the Foldy-Wouthuysen (FW) transformation [52]. The process of applying the FW transformation to the Dirac Hamiltonian has been detailed extensively in various publications [49, 53, 54] and will not be reiterated here for brevity. The final form is represented by Eq. (6) in [49].

$$\hat{H}^{FW}\Psi_L = E \cdot (1 + \hat{X}^\dagger \cdot \hat{X}), \qquad (6)$$

where $\Psi_L$ is the large component of the Dirac wave function

$$\Psi_D = \begin{pmatrix} \Psi_L \\ \Psi_S \end{pmatrix}. \qquad (7)$$

The Hamiltonian, which is the Dirac Hamiltonian transformed by a unitary matrix, is presented in the equation (8) [49, 54]:

$$\hat{H}^{FW} = U^+ H_D U = V + C \cdot (\vec{\sigma} \cdot \hat{\vec{P}}) \cdot \hat{X} + \hat{X}^\dagger \cdot C \cdot (\vec{\sigma} \cdot \hat{\vec{P}}) + \hat{X}^\dagger \cdot (V - 2.m.C^2) \cdot \hat{X}, \qquad (8)$$

and $\hat{X}$ is the operator which connects the large and small components of CD via equation (9) [49, 54]

$$\Psi_S = \hat{X}\Psi_L. \qquad (9)$$

The operator $\hat{X}$ satisfies the condition represented by equation (10) [49, 54].

$$C \cdot (\vec{\sigma} \cdot \hat{\vec{P}}) = 2.m.C^2 \cdot \hat{X} + [\hat{X}, \hat{V}] + \hat{X} \cdot C \cdot (\vec{\sigma} \cdot \hat{\vec{P}}) \cdot \hat{X} \qquad (10)$$

In equations (8) and (10), $\vec{\sigma}$ is the vector of the Pauli matrices $\vec{\sigma}(\sigma_x, \sigma_y, \sigma_z)$ [30], $\hat{\vec{P}} = -i \cdot \hbar \cdot \vec{\nabla}$ is the momentum operator, m is the rest mass of the electron and c is the speed of light.

Assuming that the exact solution to the Dirac equation is known, the operator $\hat{X}$ can be represented as in equation (11) [49],

$$\hat{X} = \frac{C}{2.m.C^2 - \hat{V} + E} \cdot (\vec{\sigma} \cdot \hat{\vec{P}}), \qquad (11)$$

and equation (6) can be re-written in the form of equation (12),

$$\left\{ \hat{V} + (\vec{\sigma} \cdot \hat{\vec{P}}) \cdot \left[ \frac{C^2}{2.m.C^2 - \hat{V} + E} + \frac{C^2 \cdot E}{(2.m.C^2 - V.E)^2} \right] \cdot (\vec{\sigma} \cdot \hat{\vec{P}}) \right\} \Psi_L$$

$$= E \cdot \left\{ 1 + (\vec{\sigma} \cdot \hat{\vec{P}}) \cdot \frac{C^2}{(2.m.C^2 - \hat{V} + E)^2} \cdot (\vec{\sigma} \cdot \hat{\vec{P}}) \right\} \Psi_L, \qquad (12)$$

where $E$ is the Dirac eigen energy corresponding to the known solution. Expanding equation (12) with respect to the parameter $E/(2 \cdot m \cdot c^2 - V)$ leads [49] in the zeroth order to the ZORA Hamiltonian in equation (13).

$$H^{ZORA} = \hat{V} + \frac{1}{2m} \cdot (\vec{\sigma} \cdot \hat{\vec{P}}) \cdot \frac{1}{1 + \frac{V}{2.m.C^2}} \cdot (\vec{\sigma} \cdot \hat{\vec{P}}). \qquad (13)$$

A drawback of the ZORA Hamiltonian is its gauge invariance issue. When a constant $C$ is added to the potential $V$ creating a new potential $V' = V + C$, the energy does not shift uniformly by $C$, as it ideally should. This gauge dependence significantly impacts core orbital properties, such as in X-ray spectroscopy, and can also affect molecular structure optimizations. However, various approaches, like scaling procedures, which used in this work, can address the gauge invariance problem, making ZORA a highly effective and efficient relativistic approximation for optimizing structures and predicting properties of compounds with heavy elements [36, 43].

## 3. The Dirac-Hartree-Fock

The Dirac-Hartree-Fock (DHF) method extends the traditional Hartree-Fock framework to incorporate relativistic effects, crucial for more accurately describing the electronic structure of molecules containing heavy atoms. By building on the Dirac equation, the DHF method accounts for relativistic effects, such as spin-orbit coupling, which become pronounced as nuclear charge increases [55].

**Fundamental Concepts and Equations**

In the DHF method, the Dirac-Coulomb Hamiltonian $\hat{H}_{DC}$ serves as the starting point.

The Dirac-Hartree-Fock equations are derived by applying the variational principle to this Hamiltonian, leading to a set of self-consistent field (SCF) equations for four-component spinor orbitals $\psi_i(r)$, which account for both large and small components of the wave function:

$$[c \cdot \alpha \cdot p + \beta \cdot m \cdot c^2 + V_{ext}(r)]\psi_i(r) + \sum_j \int \frac{\psi_j^{+}(r') \cdot \psi_j(r')}{|r - r'|} \cdot d^3r' \cdot \psi_i(r) = \epsilon_i \cdot \psi_i(r). \quad (14)$$

These coupled integro-differential equations can be solved iteratively to obtain the four-component spinors and corresponding orbital energies $\epsilon_i$ [56].

**Relativistic Effects and Approximations**

Due to the complexity of the Dirac equation, it is common to use two-component methods, such as the Douglas-Kroll-Hess (DKH) and exact two-component (X2C) transformations, that simplify the DHF equations while preserving key relativistic effects [57]. These transformations remove small component terms perturbatively, reducing computational costs without significant loss of accuracy.

## 4. Relativistic density functional theory
**Relativistic Effects and the Dirac-Kohn-Sham Equations**

In relativistic DFT, the Dirac-Kohn-Sham (DKS) equations replace the non-relativistic Kohn-Sham equations, providing a relativistic framework for the one-electron density functional approach. These equations are derived from the Dirac-Coulomb Hamiltonian, used in the Dirac-Kohn-Sham equations to account for relativistic kinetic and potential energy terms in the electron system, accounting for both large and small components that reflect relativistic behavior. The foundational principles of relativistic density functional theory were established in works by Rajagopal and Callaway [58], Rajagopal [59], MacDonald and Vosko [60], as well as Ramana and Rajagopal [61]. We use $\rho$ and $\vec{j} = (j_x, j_y, j_z)$ for the charge density and the spatial part of the four-current. Analogous to the non-relativistic case, the Kohn–Sham method can be extended to the relativistic framework by expressing the energy functional as follows.

$$E[\rho, \vec{j}] = T_S[\rho, \vec{j}] + \int V(\vec{r}) \cdot \rho(\vec{r}) \cdot d^3r + \frac{1}{2} \cdot \iint \frac{\rho(\vec{r_1}) \cdot \rho(\vec{r_2})}{|\vec{r_1} - \vec{r_2}|} \cdot d^3r_1 \cdot d^3r_2$$
$$- \frac{1}{2 \cdot C^2} \cdot \iint \frac{\vec{j}(\vec{r_1}) \cdot \vec{j}(\vec{r_2})}{|\vec{r_1} - \vec{r_2}|} \cdot d^3r_1 \cdot d^3r_2 + E_{XC}[\rho, \vec{j}], \quad (15)$$

where $T_S[\rho, \vec{j}]$ represents the kinetic energy of a reference system composed of non-interacting Dirac particles. Consequently, the Kohn–Sham approach results in a self-consistent solution to a one-particle Dirac equation, with the Kohn–Sham orbitals being four-component Dirac spinors.

In this context, $m$ denotes the electron mass, $c$ the speed of light, $\hat{\vec{P}} = -i \cdot \hbar \cdot \vec{\nabla}$ the linear momentum operator, and $\vec{\sigma}\,(\sigma_x, \sigma_y, \sigma_z)$ the vector of Pauli spin matrices. Following the SI convention, there is no $1/c$ factor in the Lorentz force. With these units, the speed of light is $c \approx 137$, the Bohr magneton is $\mu_B = 1/2$, and the vacuum permeability is $\mu_0 = 4 \cdot \pi/c^2$. The effective Kohn–Sham scalar and vector potentials are $V_{eff}$ and $\vec{A}_{eff}(\vec{r})$, respectively.

$$V_{eff} = V(\vec{r}) + \int \frac{\rho(\vec{s})}{(\vec{r}-\vec{s})} \cdot d^3s + \frac{\delta E_{XC}}{\delta \rho}(\vec{r})$$

$$\vec{A}_{eff}(\vec{r}) = \frac{-1}{C^2} \cdot \int \frac{\vec{j}(\vec{s})}{(\vec{r}-\vec{s})} \cdot d^3s + \frac{\delta E_{XC}}{\delta \vec{j}}(\vec{r}) \quad (16)$$

The effective potential $V_{eff}$ consists of three components: the external potential (typically the electrostatic potential created by the nuclear structure and scaled by the electron charge), the Hartree potential $V_H$ (the second term), and the scalar exchange-correlation potential $V_{xc}$, which is derived as the functional derivative of the exchange-correlation energy. When external magnetic fields are absent, $\vec{A}_{eff}(\vec{r})$ includes only two terms, originating from the current-current contribution to the Hartree energy and a vector exchange-correlation component, also derived as a functional derivative of $E_{xc}$. The main question then is which occupied Kohn–Sham orbitals $\psi_i$ are required to determine the kinetic energy, density, and current:

$$T_S = \sum_i^{occ} \left\langle \begin{pmatrix} \phi_i \\ \chi_i \end{pmatrix} \middle| \begin{pmatrix} m.C^2 & C.\vec{\sigma}.\vec{P} \\ C.\vec{\sigma}.\vec{P} & -m.C^2 \end{pmatrix} \middle| \begin{pmatrix} \phi_i \\ \chi_i \end{pmatrix} \right\rangle, \quad (17)$$

$$\rho(\vec{r}) = \sum_i^{occ} \psi_i^\dagger(\vec{r}) \cdot \psi_i(\vec{r}) = \sum_i^{occ} \phi_i^\dagger(\vec{r}) \cdot \phi_i(\vec{r}) + \chi_i^\dagger(\vec{r})\chi_i(\vec{r}), \quad (18)$$

$$\vec{j}(\vec{r}) = \sum_i^{occ} \psi_i^\dagger(\vec{r}) \cdot \begin{pmatrix} 0 & C.\vec{\sigma} \\ C.\vec{\sigma} & 0 \end{pmatrix} \cdot \psi_i(\vec{r}) = C \cdot \sum_i^{occ} \phi_i^\dagger(\vec{r}).\vec{\sigma}.\chi_i(\vec{r}) + \chi_i^\dagger(\vec{r}).\vec{\sigma}.\phi_i(\vec{r}), \quad (19)$$

since the Dirac operator includes eigenfunctions with both positive and negative energies.

The Dirac-Kohn-Sham orbitals $\psi_i$ are obtained by solving the DKS equations:

$$[c \cdot \alpha \cdot p + \beta \cdot m \cdot c^2 + V_{ext}(r) + V_{XC}]\psi_i(r) = \epsilon_i \cdot \psi_i(r), \quad (20)$$

where $V_{XC}$ is the exchange-correlation potential, a functional of the electron density that captures electron correlation and exchange effects. Relativistic DFT relies on relativistic functionals for $V_{XC}$ that accommodate the spin-orbit coupling and other relativistic effects in heavy-element molecules [57]. If an external magnetic field were present, it would not significantly alter the approach: the vector potential $\vec{A}(\vec{r})$, representing the external magnetic field, simply needs to be added to $\vec{A}_{eff}(\vec{r})$. Thus, the absence of an external magnetic field does not simplify the formulation. By limiting exchange-correlation functionals to be only dependent on charge density and ignoring the current-current term in the Hartree energy (second term in Eq. (17)), $\vec{A}_{eff}(r)$ vanishes from the DKS equation. For (nearly) neutral systems, the current-current contribution to the Hartree term is typically minor (as only unpaired electrons in the valence shell contribute), so it is often neglected. For stationary closed-shell systems, it disappears entirely since $\vec{j}(\vec{r}) = 0$ throughout. It's important to note that Breit contributions to the exchange-correlation energy also appear in closed-shell systems. As a result, Breit contributions to $E_{xc}$ are significantly larger than those to the Hartree energy. Given the effectiveness of the Dirac–Coulomb operator in wavefunction-based methods, it is reasonable to adopt a relativistic density functional approach that consistently omits the Breit interaction, aligning it closely with the Dirac–Fock–Coulomb method.

### Spin-Orbit Coupling and Scalar Relativistic Effects

Relativistic DFT includes both scalar relativistic effects, such as mass-velocity and Darwin terms, and spin-orbit coupling, which arise from the inclusion of the Dirac formalism in the DKS equations. Scalar relativistic effects are addressed by the use of effective core potentials (ECPs) or pseudo-potentials, or by transformations such as the Douglas-Kroll-Hess (DKH) or exact two-component (X2C) methods, which simplify the four-component Dirac equation into a two-component form [62]. Spin-orbit coupling effects, on the other hand, are particularly significant for systems with high atomic numbers and are naturally incorporated within the four-component DKS formalism. These effects alter electronic states and molecular orbitals, impacting both ground and excited-state properties and making relativistic DFT essential for accurate modeling of transition metal complexes, actinides, and lanthanides [63].

### Exchange-Correlation Functionals in Relativistic DFT

The exchange-correlation (XC) functionals used in relativistic DFT must also be adapted for relativistic systems, particularly to accommodate spin polarization effects. While common functionals (e.g., PBE, B3LYP) are used in both non-relativistic and relativistic DFT, their accuracy is often improved in the relativistic context by incorporating spin-orbit coupling contributions directly into the functional or through post-SCF spin-orbit corrections [64].

## 5. Relativistic Coupled-Cluster
### The Relativistic Hamiltonian

The coupled-cluster (CC) method itself is a powerful technique in quantum chemistry, known for its ability to incorporate electron correlation systematically and with high precision. By applying the coupled-cluster framework in a relativistic context, RCC theory takes into account not only non-relativistic electron correlation effects but also the relativistic changes in electron motion and interaction. RCC theory does this by extending traditional CC methods to include four-component spinors derived from the Dirac equation, or by using two-component approximations (like the Douglas-Kroll-Hess method or exact two-component theory) to simplify calculations while preserving essential relativistic features. RCC theory combine the formalism of relativistic quantum mechanics (typically through the Dirac equation) with the coupled-cluster (CC) framework for incorporating electron correlation. Below are the fundamental equations that illustrate the structure of RCC theory.

In RCC theory, the starting point is the Dirac-Coulomb Hamiltonian, which is the Hamiltonian for an N-electron system under relativistic treatment.

### The Coupled-Cluster Ansatz for the Wavefunction

In CC theory, the wavefunction $|\Psi\rangle$ is expressed using an exponential ansatz:

$$|\Psi\rangle = e^{\hat{T}}|\Phi_0\rangle, \qquad (21)$$

where, $|\Phi_0\rangle$ is a single-reference Hartree-Fock (HF) determinant (typically from a Dirac-Hartree-Fock calculation in the relativistic case) and $\hat{T}$ is the cluster operator, which accounts for electron correlations. The cluster operator $\hat{T}$ can be expanded as:

$$\hat{T} = \sum_n \hat{T}_n, \qquad (22)$$

where $\hat{T}_n$ represents n-particle excitations. In practice, truncations such as CCSD (Coupled-Cluster with Single and Double excitations) or CCSD(T) (CCSD with perturbative Triple excitations) are common [65].

**The Cluster Amplitude Equations**

The key objective in RCC theory is to determine the amplitudes $t_i^a, t_{ij}^{ab}, \ldots$ in $\hat{T}$ such that the projected Schrödinger equation is satisfied:

$$\langle \Phi_i^a | e^{-\hat{T}} \cdot \hat{H}_{DC} \cdot e^{+\hat{T}} | \Phi_0 \rangle = 0, \qquad (23)$$

where $\langle \Phi_i^a |$ denotes excited determinants relative to the reference state $|\Phi_0\rangle$. This non-linear equation is solved iteratively for the cluster amplitudes $t$, capturing the electron correlation in the presence of relativistic effects [62, 66].

**Energy Computation**

Once the cluster amplitudes are determined, the energy of the system can be obtained as the expectation value of the Hamiltonian with the CC wavefunction:

$$E = \langle \Phi_0 | e^{-\hat{T}} \cdot \hat{H}_{DC} \cdot e^{+\hat{T}} | \Phi_0 \rangle. \qquad (24)$$

In practice, this expectation value reduces to:

$$E = \langle \Phi_0 | \hat{H}_{DC} + [\hat{H}_{DC}, \hat{T}] + \frac{1}{2} \cdot [[\hat{H}_{DC},]\hat{T}, \hat{T}] + \cdots | \Phi_0 \rangle. \qquad (25)$$

Here, the Baker-Campbell-Hausdorff (BCH) expansion is used to simplify the exponential expression. This series is typically truncated at the fourth commutator term or lower for computational feasibility [67].

## III- Computational details

All computations in this study were performed using the DIRAC code, version 2024 [33]. The importance of relativistic effects was evaluated by conducting at the Hartree-Fock (HF) level relativistic and non-relativistic calculations, using the scaled ZORA (Zeroth Order Regular Approximation) Hamiltonian [37, 46, 68] -normalized over 4-components- for the former and Schrödinger Hamiltonian for the latter computations. These calculations are referred to as ZORA-HF and NR-HF, respectively. In order to assess the correlation effects in the relativistic case, corresponding calculations were carried out using either the B3LYP function density-functional theory (DFT) method, or the coupled cluster (CCSD(T)) method[69], denoted respectively as ZORA-DFT and ZORA-CC methods. The full Dirac-Coulomb Hamiltonian with the Gaunt interaction[57] at the CCSD(T) level was also utilized in order to predict the values of the fundamental atomic properties of the studied atomic systems and will be referred to as FDC-G-CC. For all-electron calculations, the quadruple-zeta (dyall.4zp) basis sets for the 4s, 5s, 6s, and 7s elements was employed[70], incorporating the nuclear Gaussian charge distribution model [71].

## IV- Results and discussion

In this section, we present the computed atomic properties for $X^n$ systems, where X represents F, Cl, Br, I, At, Ts, and Au, and n denotes 0, 1- and 1+. The considered electronic configurations (for the furthest p-orbitals in the halogen systems) are conventionally, using Dirac program notations where the orbital energy increases to the right, given by [$p_x$ (2e-), $p_y$ (1e-), $p_z$ (1e-)], [$p_x$ (2e-), $p_y$ (2e-), $p_z$ (1e-)], [$p_x$ (2e-), $p_y$ (2e-), $p_z$ (2e-)] for the cationic, neutral and anionic systems, respectively. The atomic properties under scrutiny are radii (R), orbitals energies, first and second ionization energies (IE), electron affinities (EA) and polarizabilities (α).

In order to estimate the radii of orbitals we computed the $r_{rms} \equiv \langle r \rangle^{1/2} = \langle x^2 \rangle + \langle y^2 \rangle + \langle z^2 \rangle$, and Tables 1-a and 1-b show the radii for the outermost p-orbitals, as determined using NR-HF, ZORA-HF, and ZORA-DFT methods with the dyall.4zp basis set.

Table 1-a: Radii (Å) of the outermost p-orbitals for "Light" $X^n$ (X = F, Cl, Br; n = 0, 1+, 1-) systems in the ground state, as calculated using NR-HF, ZORA-HF, and ZORA-DFT methods with the dyall.4zp basis set. For each system, the configuration energy increases to the right. Dirac program spectroscopic notation $^{2S+1}L_{J,M}$ is used.

| Method | Charge | F | Cl | Br | F | Cl | Br | F | Cl | Br |
|---|---|---|---|---|---|---|---|---|---|---|
| $^{2S+1}L_{J,M}$ | | $p_{1/2,1/2}$ | | | $p_{3/2,-3/2}$ | | | $p_{3/2,1/2}$ | | |
| NR-HF | + | 0.588 | 0.990 | 1.132 | 0.576 | 0.976 | 1.118 | 0.576 | 0.976 | 1.118 |
| ZORA-HF | | 0.583 | 0.982 | 1.110 | 0.579 | 0.982 | 1.125 | 0.579 | 0.982 | 1.125 |
| ZORA-DFT | | 0.586 | 0.978 | 1.105 | 0.598 | 0.993 | 1.136 | 0.598 | 0.993 | 1.136 |
| $^{2S+1}L_{J,M}$ | | $p_{1/2,1/2}$ | | | $p_{3/2,1/2}$ | | | $p_{3/2,-3/2}$ | | |
| NR-HF | 0 | 0.664 | 1.073 | 1.216 | 0.664 | 1.073 | 1.216 | 0.632 | 1.040 | 1.180 |
| ZORA-HF | | 0.657 | 1.063 | 1.192 | 0.664 | 1.073 | 1.220 | 0.645 | 1.054 | 1.199 |
| ZORA-DFT | | 0.669 | 1.066 | 1.191 | 0.671 | 1.073 | 1.222 | 0.692 | 1.090 | 1.236 |
| $^{2S+1}L_{J,M}$ | | $p_{1/2,1/2}$ | | | $p_{3/2,1/2}$ | | | $p_{3/2,-3/2}$ | | |
| NR-HF | - | 0.773 | 1.183 | 1.331 | 0.773 | 1.183 | 1.331 | 0.773 | 1.183 | 1.331 |
| ZORA-HF | | 0.772 | 1.178 | 1.302 | 0.775 | 1.186 | 1.342 | 0.775 | 1.186 | 1.342 |
| ZORA-DFT | | 0.801 | 1.192 | 1.315 | 0.804 | 1.202 | 1.361 | 0.804 | 1.202 | 1.361 |

Table 1-b: Radii (Å) for "Heavy" $Y^n$ (Y= I, At, Ts, Au; n= 0, 1+, 1-) systems in the ground state, as determined by the NR-HF, ZORA-HF, and ZORA-DFT methods employing the dyall.4zp basis set. For each system, the configuration energy increases to the right. Dirac program spectroscopic notation $^{2S+1}L_{J,M}$ is used.

| Method | Charge | I | At | Ts | I | At | Ts | I | At | Ts | Au |
|---|---|---|---|---|---|---|---|---|---|---|---|
| $^{2S+1}L_{J,M}$ | | $p_{1/2,1/2}$ | | | $p_{3/2,-3/2}$ | | | $p_{3/2,1/2}$ | | | $d_{5/2,1/2}$ |
| NR-HF | + | 1.339 | 1.446 | 1.595 | 1.325 | 1.431 | 1.580 | 1.325 | 1.431 | 1.580 | 0.877 |
| ZORA-HF | | 1.289 | 1.294 | 1.212 | 1.336 | 1.454 | 1.658 | 1.336 | 1.454 | 1.658 | 0.912 |
| ZORA-DFT | | 1.280 | 1.102 | 1.219 | 1.345 | 1.469 | 1.670 | 1.345 | 1.469 | 1.670 | 0.927 |
| $^{2S+1}L_{J,M}$ | | $p_{1/2,1/2}$ | | | $p_{3/2,1/2}$ | | | $p_{3/2,-3/2}$ | | | $s_{1/2,1/2}$ |
| NR-HF | 0 | 1.426 | 1.534 | 1.686 | 1.426 | 1.534 | 1.686 | 1.392 | 1.500 | 1.651 | 2.106 |
| ZORA-HF | | 1.367 | 1.354 | 1.240 | 1.437 | 1.572 | 1.819 | 1.416 | 1.548 | 1.787 | 1.744 |
| ZORA-DFT | | 1.362 | 1.350 | 1.249 | 1.438 | 1.578 | 1.814 | 1.451 | 1.591 | 1.829 | 1.674 |
| $^{2S+1}L_{J,M}$ | | $p_{1/2,1/2}$ | | | $p_{3/2,1/2}$ | | | $p_{3/2,-3/2}$ | | | $d_{5/2,1/2}$ |
| NR-HF | - | 1.546 | 1.658 | 1.814 | 1.546 | 1.658 | 1.814 | 1.546 | 1.658 | 1.814 | 2.627 |
| ZORA-HF | | 1.470 | 1.423 | 1.260 | 1.570 | 1.739 | 2.069 | 1.570 | 1.739 | 2.069 | 2.147 |
| ZORA-DFT | | 1.477 | 1.424 | 1.273 | 1.139 | 1.766 | 2.074 | 1.139 | 1.766 | 2.074 | 1.941 |

Table 2-a: Absolute values of the changes in atomic radii, evaluated in angstrom, due to relativistic effects. Dirac program spectroscopic notation $^{2S+1}L_{J,M}$ is used.

| Charge | F | Cl | Br | F | Cl | Br | F | Cl | Br |
|---|---|---|---|---|---|---|---|---|---|
| $^{2S+1}L_{J,M}$ | $p_{1/2,1/2}$ | | | $p_{3/2,1/2}$ | | | $p_{3/2,-3/2}$ | | |
| + | 0.005 | 0.008 | 0.022 | 0.003 | 0.006 | 0.007 | 0.003 | 0.006 | 0.007 |
| $^{2S+1}L_{J,M}$ | $p_{1/2,1/2}$ | | | $p_{3/2,1/2}$ | | | $p_{3/2,-3/2}$ | | |
| 0 | 0.007 | 0.010 | 0.024 | 0.000 | 0.000 | 0.004 | 0.013 | 0.014 | 0.019 |
| - | 0.001 | 0.005 | 0.029 | 0.002 | 0.003 | 0.011 | 0.002 | 0.003 | 0.011 |

Figure 1 shows that, for all the studied halogen systems, the relativistic effects, introduced via the ZORA Hamiltonian, lead to the shrinkage of the $p_{1/2}$ orbital and expansion of the $p_{3/2}$ orbitals. Accordingly, one would expect $p_{1/2}$ ($p_{3/2}$) orbitals to become more (less) stable in halogen systems under the influence of relativistic effects. Consequently, the first (second) ionization energy will be smaller (larger) compared to the non-relativistic value. This prediction is validated by analyzing the relativistic and non-relativistic results for orbital energies and first and second ionization potentials, detailed in Tables 3,4 and 5. Figure 1 illustrates that, under relativistic effects, the $p_{1/2}$ orbital contracts increasingly with the atomic number for the three halogens (I, At, Ts) and their ions, while the $p_{3/2}$ orbitals expand with the atomic number increasing. Furthermore, relativistic effects have a more pronounced impact on the $p_{1/2}$ orbital compared to the $p_{3/2}$ orbitals. This observation is consistent with the known result stating that "the combined effect of mass-velocity and spin-orbit interactions nearly cancels out for $p_{3/2}$ electrons but reinforces for $p_{1/2}$ electrons [10]". Tables 2-a and 2-b show the deviations ($\Delta r = r_{NR} - r_R$) to further evaluate the impact of relativistic effects on the contraction and expansion of outer orbital radii in neutral and ionic systems.

In order to gauge the significance of relativistic effects on radii, Figure 2 illustrates the absolute difference between relativistic and non-relativistic radii, which underlines the influence of the orbital type and its occupation, in addition to the atomic number, on these effects. For example, when comparing Ts$^+$ with At$^-$, the relativistic effects on the $p_{1/2}$ orbitals become more pronounced with increasing atomic number, since the occupation number is the same in both atomic systems. Conversely, for the $p_{3/2;m}$ orbitals (m = ±3/2, ±1/2), although Ts$^+$ has a higher atomic number than At$^-$, relativistic effects are more significant in At$^-$ due to its orbital occupancy. Additionally, the fractional change of relativistic effects on the radius when moving from X$^-$ to X and from X to X$^+$ (where X = At, Ts) is nearly three (two) times greater in the case of Ts for the $p_{3/2}$ ($p_{1/2}$) orbital. We list the absolute values of the changes, due to relativistic effects, of the radii for the light halogen elements in Table 2a, and for the heavy halogen elements and gold in Table 2b.

Table 2-b: Absolute values of the changes in atomic radii, evaluated in angstrom, due to relativistic effects. Dirac program spectroscopic notation $^{2S+1}L_{J,M}$ is used.

| Charge | I | At | Ts | I | At | Ts | I | At | Ts | Au |
|---|---|---|---|---|---|---|---|---|---|---|
| $^{2S+1}L_{J,M}$ | | $p_{1/2,1/2}$ | | | $p_{3/2,1/2}$ | | | $p_{3/2,-3/2}$ | | $d_{5/2,\,1/2}$ |
| + | 0.050 | 0.152 | 0.383 | 0.011 | 0.023 | 0.078 | 0.011 | 0.023 | 0.078 | 0.035 |
| $^{2S+1}L_{J,M}$ | | $p_{1/2,1/2}$ | | | $p_{3/2,1/2}$ | | | $p_{3/2,-3/2}$ | | $s_{1/2,1/2}$ |
| 0 | 0.059 | 0.180 | 0.446 | 0.011 | 0.038 | 0.133 | 0.024 | 0.048 | 0.136 | 0.362 |
| - | 0.076 | 0.235 | 0.554 | 0.024 | 0.081 | 0.255 | 0.024 | 0.081 | 0.255 | 0.480 |

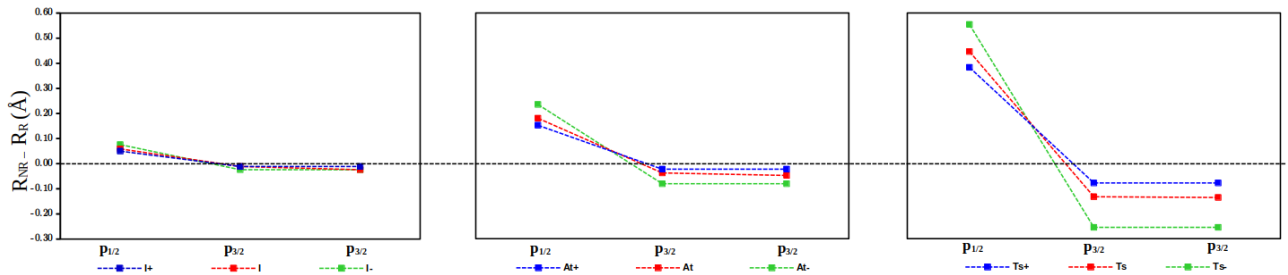

Figure 1: The effect of relativistic influences on the radii of the systems examined. The orbitals energy increases to the right.

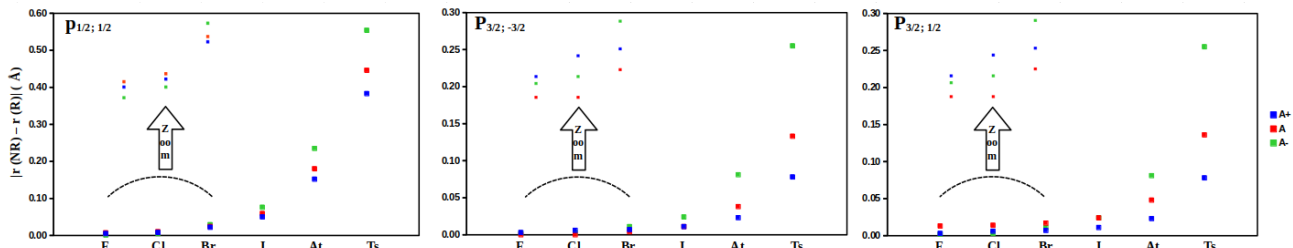

Figure 2: The magnitude of relativistic effects on the radii of the outermost p-orbitals in halogen systems. The zooming helps just for a qualitative distinction between levels near the zero value.

Table 3 shows the relativistic and non-relativistic energies for the outer orbitals of the atoms I, At, Ts, and Au in their respective states, calculated at the HF level with the dyall.4zp basis set. Also, the absolute differences between the relativistic and non-relativistic values were computed to assess the importance of the relativistic effects for each orbital.

Table 3: The energies, evaluated in a.u., of the three outermost orbitals for the I, At, and Ts atoms. All values are negative. . Dirac program spectroscopic notation $^{2S+1}L_{J,M}$ is used.

|  | I | | | At | | | Ts | | | Au |
|---|---|---|---|---|---|---|---|---|---|---|
| $^{2S+1}L_{J,M}$ | $p_{1/2,1/2}$ | $p_{3/2,1/2}$ | $p_{3/2,-3/2}$ | $p_{1/2,1/2}$ | $p_{3/2,1/2}$ | $p_{3/2,-3/2}$ | $p_{1/2,1/2}$ | $p_{3/2,1/2}$ | $p_{3/2,-3/2}$ | $s_{1/2,1/2}$ |
| E (NR-HF) | 0.392 | 0.392 | 0.351 | 0.370 | 0.370 | 0.332 | 0.343 | 0.343 | 0.308 | 0.160 |
| E (ZORA-HF) | 0.425 | 0.379 | 0.319 | 0.464 | 0.329 | 0.274 | 0.627 | 0.258 | 0.209 | 0.218 |
| \|E(NR)-E(R)\| | 0.033 | 0.013 | 0.032 | 0.094 | 0.041 | 0.058 | 0.284 | 0.085 | 0.099 | 0.058 |

In line with the observations related to atomic radii, Table 3 shows that for iodine (I), astatine (At), and tennessine (Ts), the relativistic effects augments (diminishes) stabilization to the $p_{1/2}$ ($p_{3/2}$) orbitals. Figure 3 shows the valence orbital energy levels listed in Table 3, where we note that even in the non-relativistic case, the $p_z$ orbital is higher in energy than the $p_x$ and $p_y$ orbitals due to different occupancy leading through self - consistent HF method to different energies. In addition, the figure shows the effect of spin-orbit coupling, where even orbitals with same occupancy, i.e. $p_x$ and $p_y$, are split into non-degenerate orbitals, and where the splitting magnitude increases with the atomic number. Table 3 shows that the relativistic effects on the gold (Au) 6s orbital are comparable to those on the $p_{3/2}$ ones in the heavier astatine and tennessine. As the relativistic effects on the gold physical/chemical properties are known to be substantive, we expect them to be equally important for the heavy halogens. However, we note that relativistic effects on the $p_{1/2}$ orbital of tennessine are far greater in magnitude than those in gold, which would likely influence its second ionization potential, making the non-relativistic estimation for the latter unreliable.

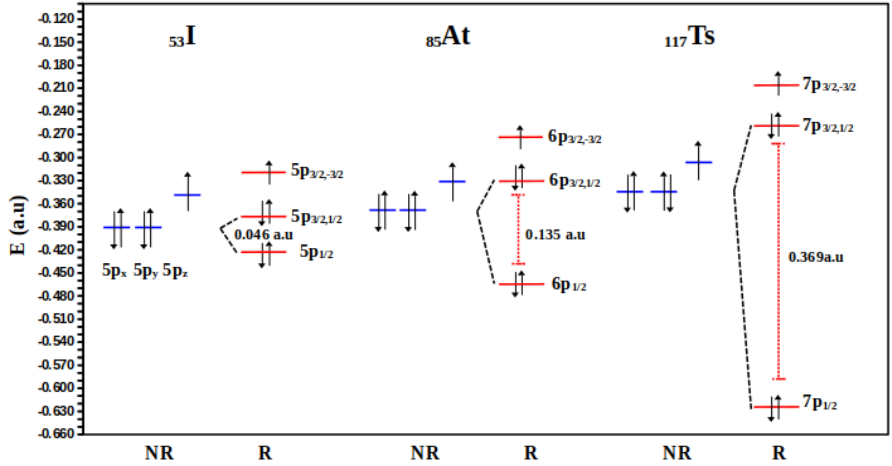

Fig 3: splitting the valence orbital under the spin-orbit effect

Tables 4 displays the first ionization energy (IE), defined as IE = E(A⁺) - E(A⁰) for the atom A, of halogen and gold atoms, calculated using different relativistic and non-relativistic methods. It shows that within the halogen group, the first IE decreases as the atomic number increases, regardless of the calculation method used. Additionally, pure relativistic effects, compared with the non-relativistic case, tend to reduce the IE, with this reduction becoming more pronounced for elements like At and particularly Ts, where these relative effects approach in magnitude those observed in gold which exceed the At case, although At is heavier than Au, while fall behind the Ts case. Among the various methods, including the highly accurate FDC-G-CC approach, the non-relativistic HF ionization potentials for fluorine through iodine align more closely with experimental values. For At, which has two experimental ionization potentials, relativistic effects bring the non-relativistic IE closer to the experimental value by approximately 2.5% (0.12%) for the first (second) experimental data. On the other hand, regarding the same heavy halogens At and Ts, the ZORA-CC and FDC-G-CC methods, compared to others, provide very similar IEs that are the closest to experimental values. For gold, however, the non-relativistic HF IE deviates from the experimental value by about 3.3eV, whereas this deviation is only 0.7 (0.6) eV for astatine, which is heavier than gold. This pattern is reversed with the FDC-G-CC method, which yields a deviation of 0.1eV for gold, compared to 0.2 (0.3) eV for astatine. Incorporating relativistic effects into the gold IE calculation reduces the deviation from experimental values by nearly 50%, while for astatine, the reduction is only 0.7%. Concerning pure correlation effects, comparisons of ZORA-HF, ZORA-DFT, and ZORA-CC results show that for the halogen elements, and unlike gold, the magnitudes of the deviations due to correlation effects are larger in the DFT method than in the CC method. Comparing third with sixth (tenth) lines of Table 4, we see that for elements from fluorine to iodine, correlation effects, estimated using DFT (CC) method, are notably more significant than relativistic effects. On the other hand, for astatine, correlation and relativistic effects are comparable, while for tennessine and gold, it is the relativistic effects which become more important. When relativistic and correlations effects are introduced using the ZORA Hamiltonian within the CC method, it becomes clear that relativistic effects dominate over correlation effects in all studied systems (halogens and gold).

**Table 4: First Ionization Energies (eV) for Gold and Halogen Atoms computed via various methods, using dyall.4zp Basis Set.**

| Method | F | Cl | Br | I | At | Ts | Au |
|---|---|---|---|---|---|---|---|
| NR-HF | 17.609 | 12.968 | 11.837 | 10.532 | 9.957 | 9.263 | 5.915 |
| ZORA-HF | 17.249 | 12.702 | 11.430 | 9.950 | 8.690 | 6.783 | 7.643 |
| \|NR-HF - ZORA-HF\| | 0.360 | 0.266 | 0.407 | 0.582 | 1.267 | 2.480 | 1.728 |
| [(NR-HF - ZORA HF)/NR-HF]×100 % | 2.044 | 2.051 | 3.438 | 5.526 | 12.725 | 26.773 | 29.21 |
| ZORA-DFT | 20.520 | 14.424 | 12.855 | 11.135 | 9.826 | 8.075 | 8.893 |
| \|ZORA-HF - ZORA-DFT\| | 3.271 | 1.722 | 1.425 | 1.185 | 1.136 | 1.292 | 1.250 |
| [(ZORA-HF - ZORA-DFT)/ZORA-HF]×100 % | 15.941 | 13.557 | 12.467 | 11.910 | 13.072 | 19.048 | 16.36 |
| ZORA-CC | 17.134 | 12.569 | 11.585 | 10.194 | 9.027 | 7.385 | 9.144 |
| \|ZORA-HF - ZORA-CC\| | 0.115 | 0.133 | 0.155 | 0.244 | 0.337 | 0.602 | 1.501 |
| [(ZORA-HF - ZORA-CC)/ZORA-HF]×100 % | 0.667 | 1.047 | 1.356 | 2.452 | 3.878 | 8.875 | 19.639 |
| FDC-G-CC | 17.131 | 12.567 | 11.582 | 10.191 | 9.024 | 8.815 | 9.136 |
| Exp | 17.423[a] | 12.968[b] | 11.814[b] | 10.451[b] | 9.2[c] 9.318[d] | — | 9.226[e] |
| [(EXP - NR-HF)/EXP]×100 % | 1.068 | 0.000 | 0.195 | 0.775 | 8.111 6.858 | — | 35.888 |
| [(EXP - ZORA-HF)/EXP]×100 % | 0.999 | 2.051 | 3.250 | 4.794 | 5.646 6.740 | — | 17.158 |
| [(EXP - ZORA-DFT)/EXP]×100 % | 17.775 | 11.228 | 8.812 | 6.545 | 6.688 5.452 | — | 3.609 |
| [(EXP - ZORA-CC)/EXP]×100 % | 1.658 | 3.174 | 1.977 | 2.521 | 1.986 3.122 | — | 0.889 |
| [(EXP - FDC-G-CC)/EXP]×100 % | 1.676 | 3.092 | 1.964 | 2.488 | 2.020 3.155 | — | 0.976 |

[a] Ref 70. [b] Ref 1. [c] Ref 11. [d] Ref 13. [e] Ref 22.

We see in Fig. 4 that the deviations do not show a linear relationship with the increase of atomic number:

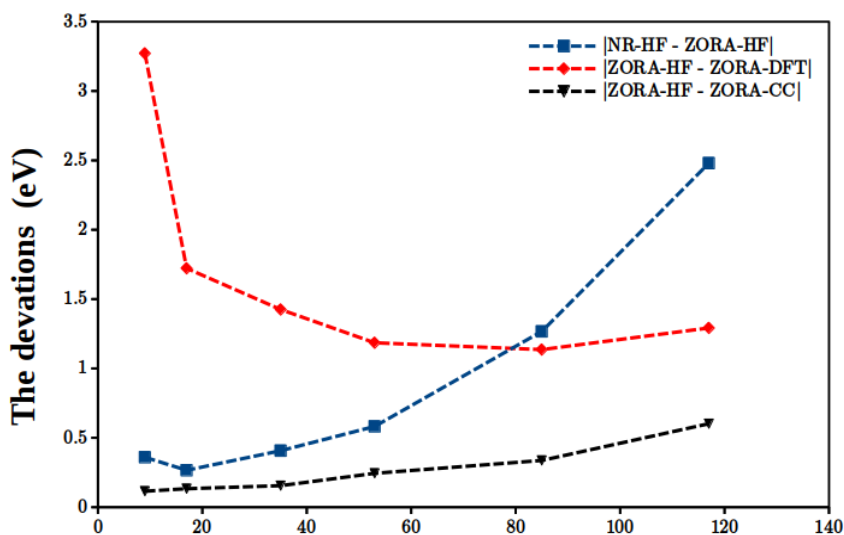

Fig. 4: The deviations (vertical axis) with respect to atomic number (horizontal axis)

Table 5 displays both the relativistic and non-relativistic estimations of the second IE for the atoms I, At and Ts, where it corresponds to removing an $np_{1/2}$ electron.

Table 5: Second Ionization Energies (eV) for Gold and Halogen Atoms computed via various methods, using dyall.4zp Basis Set.

| Method | I | At | Ts |
|---|---|---|---|
| NR-HF | 18.766 | 17.590 | 16.190 |
| ZORA-HF | 19.674 | 20.275 | 23.985 |
| \|NR-HF - ZORA-HF\| | 0.908 | 2.685 | 7.795 |
| [(NR-HF - ZORA HF)/NR-HF]×100 % | 4.839 | 15.264 | 48.147 |
| ZORA-DFT | 21.163 | 21.720 | 24.856 |
| [(ZORA-HF - ZORA-DFT)/ZORA-HF]×100 % | 7.568 | 7.127 | 3.631 |
| ZORA-CC | 19.534 | 17.474 | 23.664 |
| [(ZORA-HF - ZORA-CC)/ZORA-HF]×100 % | 0.712 | 13.815 | 1.338 |
| FDC-G-CC | 19.514 | 17.502 | 23.613 |
| Exp | 19.131[a] | — | — |
| [(EXP - NR-HF)/EXP]×100 % | 1.908 | — | — |
| [(EXP - ZORA-HF)/EXP]×100 % | 2.838 | — | — |
| [(EXP - ZORA-DFT)/EXP]×100 % | 10.622 | — | — |
| [(EXP - ZORA-CC)/EXP]×100 % | 2.063 | — | — |
| [(EXP - FDC-G-CC)/EXP]×100 % | 2.107 | — | — |

[a] Ref 1.

In contrast to the first IE, Table 5 indicates that relativistic effects augment the value of the second IE for the atoms I, At, and Ts. This increase occurs because the second removed electron is taken from the $p_{1/2}$ orbital, which is stabilized by relativistic effects. A comparison of the 1st and 2nd IEs, in both relativistic and non-relativistic calculations using ZORA-HF and NR-HF methods, shows that relativistic effects are significantly more pronounced for the 2nd IE than for the 1st. The most notable effects are observed for the heaviest element: Ts. These relativistic effects increase from I to Ts in both IEs, as shown in third lines (|NR(HF)-ZORA(HF)|) of Tables 4, 5, with the relativistic changes in eV given for I by 0.572 (0.908) corresponding to 1st (2nd ) IE, and similarly by 1.267 (2.785) for At, and 2.480 (7.795) for Ts.

According to Tables 4 and 5, we can order the atomic systems according to the relativistic effects importance as $_{117}Ts^+(p_{1/2}) > _{79}Au(s) > _{117}Ts(p_{3/2})$, where removing an electron from the orbital delimited by the brackets leads to 1st IE of the atom to their left. Namely, the relativistic effects on the $p_{1/2}$ orbital in Ts$^+$, combined with the high atomic number of Ts, imply more significant influence on the 1st IE for Ts$^+$ compared to gold, despite the fact that relativistic effects are more significant on s-orbitals than on p-orbitals. As to Ts, the ordering is expected since we know that relativistic effects are less tangible for $p_{3/2}$ than for $p_{1/2}$.

For pure correlation effects, the comparison between ZORA-HF and ZORA-DFT methods indicates that, similar to the 1st IE, the 2nd IE increases due to correlation effects. Tables 4 and 5 demonstrate that the relativistic correlation effects within the ZORA framework lead to an increase in both IEs when using the DFT method, whereas using the CC method leads to an increase in 1st IE and a decrease in 2nd IE. With experimental values available only for iodine's 2nd IE, Table 4 shows that the agreement with experiment improves upon changing the methods along the sequence: ZORA-DFT < ZORA-HF < FDC-G-CC < ZORA-CC < NR-HF making the NR- HF the most reliable method, as long as agreement with experiment is concerned, for iodine, whereas the ZORA-DFT is the least effective method.

Numerous theoretical calculations of IEs are found in the literature [1, 11, 13]. For the halogens and gold, we have selected the most accurate theoretical values relative to experiment, as shown in Table 6-a, which also includes our best theoretical values. Table 6-b presents the relative deviations from experiment.

Table 6-a: Other authors' quoted theoretical values for the 1st and 2nd (bold font) Ionization Energies (eV).

|  | F | Cl | Br | I | At | Ts | Ref |
|---|---|---|---|---|---|---|---|
| MCDF | — | 12.679 | 11.523 | 10.152 | 9.040 | 7.310 | [1] |
|  | — | **23.537** | **21.275** | **18.773** | **17.473** | **14.877** |  |
| BP86 ECP/AV5Z |  | 11.860 | 10.630 | 10.080 | — |  | [11] |
| 4c-CCSD(T)+Breit+QED | — | — | — | — | 9.313 | 7.654 | [13] |
|  | — | — | — | — | **17.753** | **15.174** |  |

Table 6-b: The relative deviations between the experimental and theoretical (our, denoted by "Present", and others' estimations) values for the 1st and 2nd (in bold) IEs. For the At, there are two quoted experimental values, whereas no experimental value for the 2nd IE is available except for I.

|  | F | Cl | Br | I | At | Au | Ref |
|---|---|---|---|---|---|---|---|
| Δ%(NR-HF) | — | 0.000 | 0.195 | 0.775 | — |  | Present |
|  | — | — | — | **1.908** | — |  |  |
| Δ%(ZORA-HF) | 0.999 | — | — | — | — |  | Present |
| Δ%(ZORA-CC) | — | — | — | — | 1.986 |  | Present |
|  | — | — | — | — | 3.122 |  |  |
| Δ%(MCDF) | — | 2.229 | 2.463 | 2.861 | 1.846 | — | [1] |
|  | — | — | — | **1.871** | 2.941 | — |  |
| Δ%(BP86 ECP/AV5Z) | — | — | 0.389 | 1.713 | 9.446 | — | [11] |
|  | — | — | — | — | 8.178 | — |  |
| Δ%(4C-CSSD(T)+Breit+QED)) | — | — | — | — | 1.118 | — | [13] |
|  | — | — | — | — | 0.054 |  |  |

Tables 6-a and 6-b show that for the elements Cl, Br, and I, our theoretical values align more closely with experimental data than the others' estimations[1, 11, 13]. Interestingly, despite its sophistication, the MCDF method does not provide an improvement compared to the NR-HF, in that its theoretical estimations are more deviated from the experimental values. On the other hand, for Astatine, our theoretical results are in line with the other theoretical estimates.

Table 7 lists the electron affinities (EA) for gold and halogens atoms, defined as the energy gained when an additional electron is attached to a neutral atom forming a negative ion[73]. Actually, obtaining EA values accurately, be it experimentally or theoretically, is more challenging than obtaining accurate values of IEs; even for small atoms it is notoriously difficult[8, 74]. When comparing the halogen EA theoretical values with the available corresponding experimental ones, we observe that the methods taking into account correlation effects produce results which are closest to experiment for atoms ranging from fluorine to iodine.

Although the NR-HF values for halogen atoms deviate significantly from the experimental values (reaching about 62% in the case of fluorine), this deviation becomes smaller as atomic number increases, so that to reach around 3% in At. Extrapolating this dependency, one can predict the deviation between the NR-HF calculation of the Ts EA and its unavailable experimental value not to exceed 3%, corresponding to 2.4eV, which would be contrasted with the value of 1.281 eV estimated by the multi-configuration Dirac-Fock (MCDF) method in [1]. Although the MCDF constitutes a more sophisticated method than the NR-HF, however based on the At case, where NR-HF gives a better deviation from experiment than MCDF (3% versus 11%), we believe our NR-HF estimation for the Ts EA is more reliable. Note that some gold and fluor values stated in the table indicate a deviation from other ones, corresponding to different methods, amounting to more than 100%. We left these values for comparison purposes even though such huge differences may cast doubts on the validity of studies applied to these two elements.

Table 7: Electron Affinities (eV) for gold and halogen atoms, calculated using the dyall.4zp Basis Set.

| Method | F | Cl | Br | I | At | Ts | Au |
|---|---|---|---|---|---|---|---|
| NR-HF | 1.294 | 2.538 | 2.528 | 2.532 | 2.496 | 2.419 | 0.003 |
| ZORA-HF | 1.286 | 2.493 | 2.346 | 2.161 | 1.498 | 1.282 | 0.644 |
| [(NR-HF - ZORA HF)/NR-HF]×100 % | 0.62 | 1.77 | 7.20 | 14.65 | 39.98 | 47.00 | ∞ |
| ZORA-CC | 3.074 | 3.288 | 4.273 | 2.794 | 1.966 | 1.367 | 2.251 |
| [(ZORA)HF - ZORA-CC)/ZORA]HF]×100 % | 139.04 | 31.89 | 82.14 | 29.29 | 31.24 | 6.63 | 249.53 |
| ZORA-DFT | 4.962 | 4.368 | 3.963 | 3.511 | 2.829 | 1.959 | 2.670 |
| [(ZORA-HF - ZORA-DFT)/ZORA-HF]×100 % | 285.85 | 75.21 | 68.93 | 62.47 | 88.85 | 52.81 | 314.60 |
| FDC-G-CC | 3.076 | 3.290 | 4.401 | 2.796 | 1.970 | 1.397 | 2.24 |
| Exp | 3.399[a] | 3.617[b] | 3.365[b] | 3.059[b] | 2.416[c] | - | 2.309[d] |
| [(EXP - NR-HF)/EXP]×100 % | 61.93 | 29.83 | 24.87 | 17.23 | 3.31 | - | 99.87 |
| [(EXP - ZORA-HF)/EXP]×100 % | 62.17 | 31.08 | 30.28 | 29.36 | 38.00 | - | 72.11 |
| [(EXP - ZORA-CC)/EXP]×100 % | 9.56 | 9.10 | 26.98 | 8.66 | 18.63 | - | 2.51 |
| [(EXP - FDC-G-CC)/EXP]×100 % | 9.77 | 9.04 | 30.79 | 8.60 | 18.46 | - | 2.9 |
| [(EXP - ZORA-DFT)/EXP]×100 % | 45.98 | 20.76 | 17.77 | 14.78 | 17.09 | - | 15.63 |

[a] Ref [73]. [b] Ref [1]. [c] Ref [15]. [d] Ref [22].

One can get some insight when evaluating the influence of relativistic effects, comparing NR-HF values with those of relativistic method like ZORA-HF (1st and 2nd lines of Table 6), in that they tend to decrease the halogen EAs, whereas they do the reverse for gold. The interpretation of this observation can be attributed to the fact that the relativistic effects cause the expansion (contraction) of the atom's $p_{3/2}$- (s-)outer orbital in the halogen (gold) atoms, reducing (augmenting) thus the EA. It is noteworthy that the impact of relativistic effects tends to grow with the atomic number within the halogen group.

To evaluate the correlation effects on the halogen EAs, we compare ZORA-HF results with those of ZORA-CC (5th line of table 6) and/or with ZORA-DFT (7th line). The results show that correlation effects vary through the halogen group, being highest (lowest) in importance at the lightest (heaviest) halogen, namely F (Ts), where the calculations show that 78% (33%) of the electrons are correlated. We note also that, except for Br, the DFT correlation effects are more pronounced than in CC method.

The electric dipole polarizability of an atom or molecule describes the linear response of the electronic charge distribution with respect to an externally applied electric field [17]. Since most systems studied fall under the so-called open-shell systems, we used the finite-field perturbation method to calculate the polarizability[73]. The polarizability ($\alpha$), was calculated as follows:

$$\boldsymbol{\alpha}_{zz} = - (E(+F) + E(-F) - 2E(0))/F^2, \qquad (26)$$

where E is the system energy and F is the strength of the applied electric filed, set to the nominal

value 0.0005.

We chose to evaluate the average polarizability where the five outer electrons are put into 6 available spin orbitals, in contrast to the specific polarizability where 4 spin orbitals are occupied and one electron is put in the last two available orbitals.

**Table 8: Average Polarizabilities, evaluated in a.u., for Gold and Halogen Atoms computed via various methods, using dyall.4zp Basis Set.**

| Method | Charge | F | Cl | Br | I | At | Ts | Au |
|---|---|---|---|---|---|---|---|---|
| NR-HF | + | 1.485 | 8.056 | 12.043 | 20.220 | 25.346 | 33.925 | 10.547 |
| ZORA-HF | | 1.483 | 8.044 | 11.906 | 19.605 | 23.631 | 34.132 | 12.347 |
| ZORA-CC | | 1.496 | 7.830 | 11.692 | 17.387 | 23.331 | 33.870 | 13.509 |
| ZORA-DFT | | 1.594 | 8.296 | 12.284 | 20.064 | 24.491 | 34.947 | 13.954 |
| NR-HF | 0 | 2.301 | 11.707 | 17.662 | 28.555 | 35.171 | 44.673 | 106.033 |
| ZORA-HF | | 2.302 | 11.719 | 17.64 | 28.343 | 35.283 | 52.690 | 47.861 |
| ZORA-CC | | 2.321 | 11.809 | 17.757 | 24.821 | 34.979 | 52.270 | 49.046 |
| ZORA-DFT | | 2.545 | 12.299 | 18.580 | 29.593 | 37.093 | 52.843 | 36.497 |
| Exp/Rec[17] | | 3.74±0.08 | 14.6±0.1 | 21±1 | 32.9±1.3 ($^2P_{3/2}$) 33.4 ($^2P_{3/2}$) | 42 ± 4 | 76 ± 15 | 49.1 ± 4.1 39.1 ± 9.8 |
| NR-HF | - | 4.015 | 17.735 | 26.698 | 41.478 | 49.613 | 58.039 | 344.974 |
| ZORA-HF | | 4.024 | 17.796 | 26.950 | 42.202 | 53.819 | 77.720 | 158.278 |
| ZORA-CC | | 4.420 | 18.551 | 27.703 | 42.860 | 54.283 | 73.588 | 84.872 |
| R-DFT | | 4.461 | 18.919 | 28.757 | 44.733 | 56.702 | 74.894 | 104.534 |

All experimental(Exp) and recommended (Rec) values are taken from the reference [17].

Table 8 aggregates the polarizability values for both neutral and charged halogens, calculated using various methods. The results show that the polarizability increases across the halogen group with the atomic number, regardless of the atomic electric charge and of the calculation method. Comparing the atom with its ions, we note that α increases with the charge negativity, in that for the compound A, it increases following the sequence: $A^+ \rightarrow A \rightarrow A^-$, whose individual members correspond to occupancy percentages in the outermost p-subshell equaling 4/6=66% → 5/6=83% → 6/6=100%. Figure 5 depicts the difference $\alpha$(NR-HF) - $\alpha$(ZORA-HF) as a function of atomic number for each atom with its two ions.

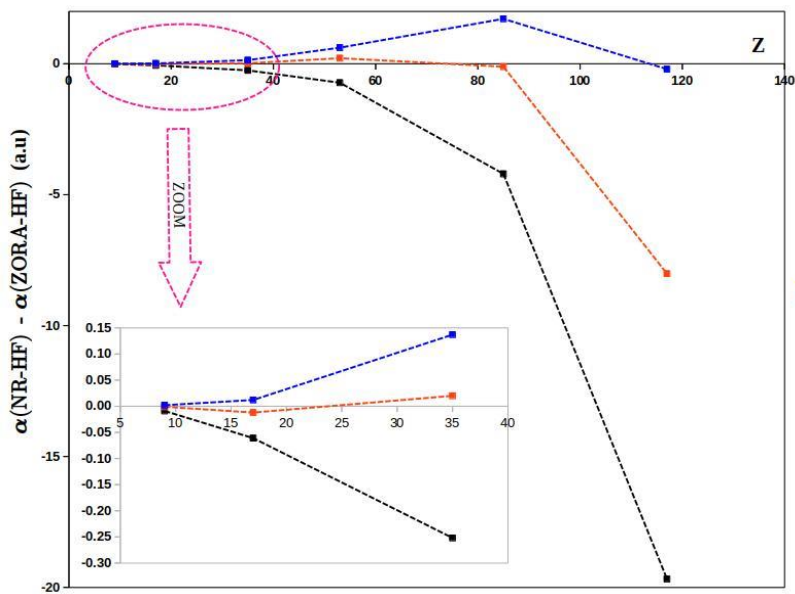

Figure 5: Average polarizability difference ($\alpha$(NR-HF) - $\alpha$(ZORA-HF)) versus the atomic number Z.

In Fig. 5, blue/red/black colors denote cations/neutrals/anions, respectively. The figure illustrates how relativistic effects reduce polarizability for cations from F to At, where it reaches a minimum after which it increases. For anionic (neutral) systems, however, there is a (no) consistent pattern, in that they tend to grow in anions with the atomic number, which would enhance polarizability, making anions progressively more polarizable with increasing atomic number. Notably, the Ts anion is the most polarizable in the halogen series, irrespective of its electric charge, when relativistic effects are considered. Figure 5 shows, as well, that the most pronounced relativistic effects, for a given atomic number, happen in anions. Some other relativistic and non-relativistic theoretical values of polarizability for gold and its ions (±1) are listed in Table 9. We note that our results in Table 8 are in line with those of Table 9, and the relativistic effects play similar roles as well.

Table 9: Other authors quoted theoretical values for the polarizability of gold and its ions 1± (a.u.).

| Method | | $Au^+$ | Au | $Au^-$ |
|---|---|---|---|---|
| HF | NR | 9.043 | 103.7 | 660.1 |
|  | R | 10.59 | 46.70 | 205.0 |
| QCISD(T) | NR | 9.650 | 64.11 | 257.1 |
|  | R | 11.61 | 35.09 | 96.03 |

All values were taken from the reference [75]. HF: unrestricted Hartree-Fock. QCISD(T): quadratic configuration interaction estimating triple contributions.

## V) Conclusion and outlook:

In order to study the relativistic and correlation effects on the halogens, we have used wave function-based approach such as (HF, CC) or density functional theory approach (DFT). The relativistic effects tend to decrease (increase) the first IE and EA ($2^{nd}$ IE) for the neutral atoms. Irrespective of the method used, we found for the heavy halogen, namely I, At, and Ts, that the correlation effects increase the $1^{st}$ IE and the EA. For the $2^{nd}$ IE, correlation effects tend to increase (decrease) it when calculated via DFT (CC).

Relativistic effects cause the atomic radii of halogens to expand, which may influence bond lengths in related molecules. Given the opposite directions in which relativistic effects, due to spin-orbit on the one hand and mass-velocity on the other hand, act in halogens, they may cancel out interpreting thus the surprisingly good agreement between the NR-HF method and experimental data for $1^{st}$ IE of Cl, $2^{nd}$ IE of I, and EA of heavy halogens, particularly At and probably Ts (by extrapolation). This suggests that computational efforts could be reduced for systems containing halogens, by restricting them to NR-HF.

This study contributes to a new axis of theoretical research investigating the fundamental atomic properties of the unfamiliar elements, At and Ts, using various theoretical methods. Moreover, our study calculated, for the first time, the polarizabilities of halogen ions, and investigated the impact of relativistic and correlation effects on these fundamental properties, reviewing them for gold atoms as well.

**Acknowledgements:**
N. C. acknowledges support from the CAS PIFI fellowship and from the Humboldt Foundation.